# Assessment of wafer-level transfer techniques of graphene with respect to semiconductor industry requirements

Sebastian Wittmann[1,5], Stephan Pindl[2], Simon Sawallich[3], Michael Nagel[3], Alexander Michalski[3], Himadri Pandey[4], Ardeshir Esteki[5], Satender Kataria[5], Max C. Lemme[5,6*]

[1]Infineon Technologies AG, Am Campeon 4, 85579 Neubiberg, Germany

[2]Infineon Technologies AG, Wernerwerkstraße 2, 93049 Regensburg, Germany

[3]Protemics GmbH, Otto-Blumenthal-Straße 25, 52074 Aachen, Germany

[4]Advantest Europe GmbH, Herrenburgerstr 130, 71034 Böblingen, Germany

[5]RWTH Aachen University, Otto-Blumenthal-Straße 2, 52074 Aachen, Germany

[6]AMO GmbH, Otto-Blumenthal-Straße 25, 52074 Aachen, Germany

*Authors to whom correspondence should be addressed:

Max C. Lemme - max.lemme@eld.rwth-aachen.de

**Abstract**

Graphene is a promising candidate for future electronic applications. Manufacturing graphene-based electronic devices typically requires graphene transfer from its growth substrate to another desired substrate. This key step for device integration must be applicable at the wafer level and meet the stringent requirements of semiconductor fabrication lines. In this work, wet and semidry transfer (i.e. wafer bonding) are evaluated regarding wafer scalability, handling, potential for automation, yield, contamination and electrical performance. A wafer scale tool was developed to transfer graphene from 150 mm copper foils to 200 mm silicon wafers without adhesive intermediate polymers. The transferred graphene coverage ranged from 97.9 % to 99.2 % for wet transfer and from 17.2 % to 90.8 % for semidry transfer, with average copper contaminations of $4.7 \times 10^{13}$ (wet) and $8.2 \times 10^{12}$ atoms/cm$^2$ (semidry). The corresponding electrical sheet resistance extracted from terahertz time-domain spectroscopy varied from 450 to 550 Ω/□ for wet transfer and from 1000 to 1650 Ω/□ for semidry transfer. Although wet transfer is superior in terms of yield, carbon contamination level and electrical quality, wafer bonding yields lower copper contamination levels and provides scalability due to existing industrial tools and processes. Our conclusions can be generalized to all two-dimensional (2D) materials.

## 1. INTRODUCTION

Graphene exhibits excellent mechanical and electronic properties and may therefore be suitable for future use in micro-, opto- and nanoelectronics as well as in nanoelectromechanical applications [1–7]. To realize such applications, it is necessary to integrate graphene from its growth substrate with the desired target substrate. Graphene transfer remains one of the most critical steps in the integration process at the wafer level because of issues related to defects, contamination and wrinkles in the transferred graphene layers [8,9]. Ultimately, graphene transfer must be automated to enable high-throughput and thus competitive graphene-based devices. Currently, there are two dominant transfer processes in the literature, wet and semidry transfer. During wet transfer, graphene is separated from its growth substrate via a wet-chemical process with the support of a polymer-containing carrier and transferred to the target substrate in a liquid environment [10–14]. Semidry transfer typically includes wet-chemical transfers as first steps to remove the growth substrate via a liquid medium, followed by dry transfer of the graphene onto the target substrate via wafer bonding [15–20]. Hence, the latter is sometimes referred to as a wafer bonding process. However, these methods leave residual adhesive polymer layers between the graphene and the target substrate, which prevent the use of subsequent high-temperature processes and reduce the accessibility of processes in industrial semiconductor manufacturing lines due to the high degree of carbon contamination. In addition to the typical transfer methods, it is possible to grow graphene directly on target substrates via metallic catalyst structures [21,22]. The disadvantage of this method is that requires high growth temperatures, which prohibit substrates with front- or back-end-of-line (FEOL, BEOL) metallization layers or diffusion regions.

In this study, the wet-chemical and semidry transfers of 150 mm graphene on copper (Cu) foils to 200 mm silicon wafers are investigated with respect to contamination, defects and electrical properties are determined optically and electrically. We compare the potential of both transfer techniques in terms of yield, potential for automation, contamination and electrical performance and assess their suitability for the integration of graphene into front-end-of-line processes.

## 2. EXPERIMENTAL

The target substrates for all experiments (wet and semidry) were silicon wafers covered with a thermal silicon dioxide film with a thickness of 300 nm.

Wet transfer was performed using commercial graphene on a 150 mm copper foil covered with polymethyl methacrylate (PMMA). The 150 mm graphene samples were placed in an ammonium persulfate solution for approximately 4 h to remove the copper layer below the graphene. A 0.1 molar ammonium persulfate solution was used for all transfer processes, as this copper etching medium does not introduce ionic contamination (e.g., $Na^+$ (NaOH), $Fe^+$ ($FeCl_3$)). The low molarity was chosen since the etching medium has an oxidizing effect, which can result in damage to the graphene. After copper etching, the 150 mm graphene layer with the PMMA carrier polymer was transferred from the ammonium persulfate solution to a deionized (DI) water bath with a 200 mm hydrophilic $SiO_2$ carrier wafer to clean the graphene for 10 min. This cleaning process was then repeated with two additional DI water baths for purification with different carrier wafers. In the last water bath, the graphene layer was picked up with the target substrate ($SiO_2$), which had previously been made hydrophilic by dipping in a DI water bath for 1 h. Subsequently, the sample was dried for 12 h under atmospheric conditions, and then the PMMA layer was removed with acetone (20 min) and isopropanol (10 min) and subsequently rinsed in a DI water bath (10 min). After this cleaning step, the sample was dried for an additional 12 h at room temperature, and then the polymer residues were removed via thermal annealing at 400°C for 2 h in a $N_2$ atmosphere under vacuum.

The semidry transfer of graphene on copper to a CMOS-compatible target substrate (e.g., Si, $SiO_2$) was carried out with a transfer tool that was designed and manufactured for this purpose. The wafer-level transfer tool consists of a mechanical press (Rotek), which has a mechanically reinforced precision hot plate as the base and a surface-treated stainless-steel stamp as the counterpart, as the basic structure. A chamber with gas inlets for dry nitrogen was added to control the atmosphere and humidity. Temperature (hot plate), pressure (hydraulic press) and humidity sensors were used to control the transfer process. The custom wafer-level transfer tool can be operated at temperatures from room temperature to approximately 230°C, at mechanical pressures (at room temperature) from 0-9.4 N/mm$^2$ and at air humidity values down to approximately 10 %. Increasing the process temperature reduces the maximum mechanical process pressure, as the hydraulic unit cannot maintain a stable pressure at higher temperatures. To be precise, the process pressure could not be kept constant above 3-4 N/mm$^2$ over several minutes at a process temperature of 100°C. The maximum process pressure was therefore set to 3 N/mm$^2$ in this work. The wafer-level transfer tool is shown in Figure 1.

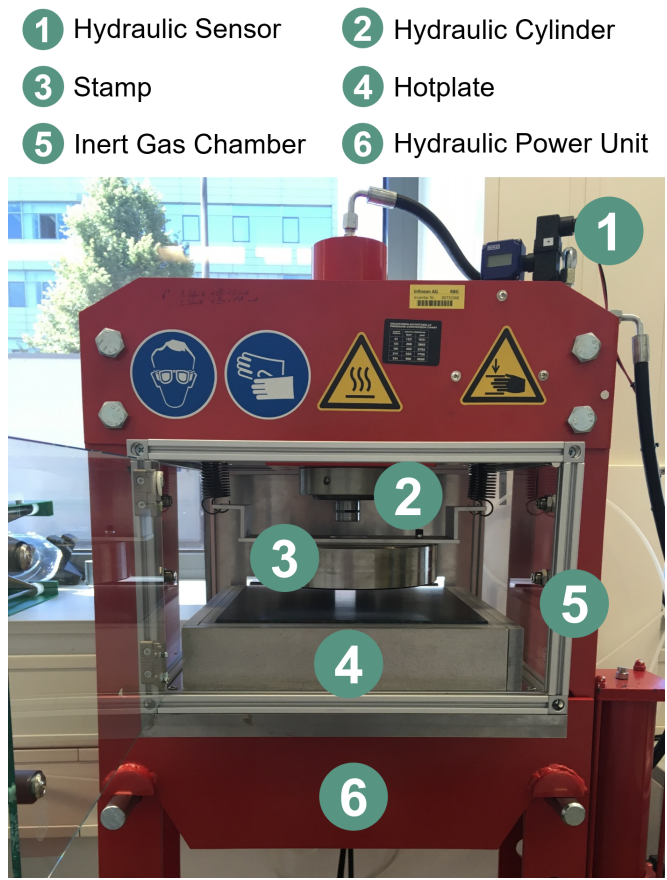

**Figure 1: Wafer-level transfer tool for the semidry transfer of graphene. It consists of a hydraulic sensor (1), a hydraulic cylinder (2), a stamp (3), a hotplate (4), an inert gas chamber (5) and a hydraulic power unit (6).**

Thermal release tape (TRT) was first laminating onto 200 mm silicon wafers to create a carrier for graphene transfer and to improve handling. 150 mm copper foils with commercially available graphene (Graphenea) were then rolled onto the carrier wafers (graphene side to TRT) and pressed onto the wafers with 4 N/mm$^2$ pressure at room temperature for 10 min in a nitrogen atmosphere. This process ensured the absence of visible air bubbles. The copper foils were then removed and cleaned using the same parameters, etching media and cleaning media as those applied for wet transfer. The cleaning parameters were kept constant during the entire experiment, as they had already been optimized via wet transfer. In contrast to the wet transfer, the semidry method includes a stabilizing carrier wafer, which allows inducing convection in the etching and purification media by using a stir bar. This improves the mixing of the media and therefore a more constant etch rate without cracking the graphene. Here, the stir bar was oriented perpendicular to the wafer surface, generating convection parallel to it so that the saturated etching medium was purged along the wafer, removing carbon residues from the copper backside. Thus, the copper foil was no longer present after 2 h, but to remain consistent with wet transfer with regard to copper contamination, the graphene remained in the etching solution for 4 h.

The tape used for transfer was NO.3198MS (Nitto) with an adhesive force of 2.5 N per 20 mm width and a release temperature of 120°C. The adhesion force of this tape is the lowest commercially available from this vendor and was chosen to minimize the adhesion from the graphene to the TRT during transfer to allow the graphene to stick to the target substrate. After copper removal and subsequent cleaning in DI water baths, the carrier wafer with graphene was dried for approx. 12 h under atmospheric conditions. The drying process reduced the absorbed water molecules in the thermal release film introduced during the copper etching and cleaning process. This is required to prevent the formation of a water film at the graphene/SiO$_2$ interface (water molecules diffuse through the cracks in the graphene layer), which would reduce the adhesion between graphene and the target substrate and lead to poor transfer.

After drying, the carrier wafer with graphene was pressed onto the target substrate under the defined pressure, temperature and atmosphere for different bonding times. Three semidry experiments were carried out with different bonding pressures for pressing the graphene/TRT/silicon stack onto the target substrate. The bonding pressure was varied from 1-3 N/mm$^2$ in a nitrogen atmosphere with a residual relative humidity of approx. 20 % and a hot plate temperature of 80°C. The hot plate temperature was chosen to reduce the water adsorbates on the target substrate and on the graphene/TRT/silicon stack while keeping it lower than the tape release temperature of 120°C. In preliminary experiments, a slight release of the TRT was observed already at a temperature of 90°C for 10 min.

In the following, the wet transfer wafers are abbreviated as WT1 (**W**et **T**ransfer Wafer No. 1), WT2 (**W**et **T**ransfer Wafer No. 2) and WT3 (**W**et **T**ransfer Wafer No. 3), and the semidry transfer wafers are abbreviated as SDT1 (**S**emi**d**ry **T**ransfer Wafer No. 1, Bonding pressure of 1 N/mm$^2$), SDT2 (**S**emi**d**ry **T**ransfer Wafer No. 2, Bonding pressure of 2 N/mm$^2$), SDT3 (**S**emi**d**ry **T**ransfer Wafer No. 3, Bonding pressure of 2 N/mm$^2$), and SDT4 (**S**emi**d**ry **T**ransfer Wafer No. 4, Bonding pressure of 3 N/mm$^2$) (see Table 1).

**Table 1: Matrix for sample identification based on their respective processing steps for wet and semidry transfer. In semidry transfer, the first bond pressure was held constant at 4 N/mm$^2$, and the second bond pressure was varied from 1-3 N/mm$^2$.**

|  | Wet transfer | Semidry transfer | | |
|---|---|---|---|---|
| **Bond pressure** | 4 N/mm$^2$ | 1 N/mm$^2$ | 2 N/mm$^2$ | 3 N/mm$^2$ |
| **Sample label** | WT1, WT2, WT3 | SDT1 | SDT2, SDT3 | SDT4 |

After bonding the graphene/TRT/silicon stack onto silicon oxide, the samples were heated to the required release temperatures to remove the carrier wafer with the release film. Then, the samples were treated with the same processes and parameters for polymer removal as for wet transfer (acetone/isopropanol/annealing). The general procedure of the semidry transfer is illustrated with photographs in Figure 2.

| 6" Graphene on copper foil | Lamination on TRT and pressed on silicon carrier | Cu wet etching | Bond step on target substrate |

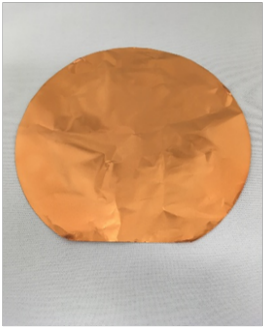 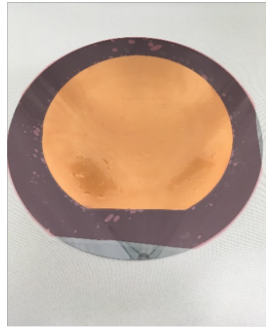 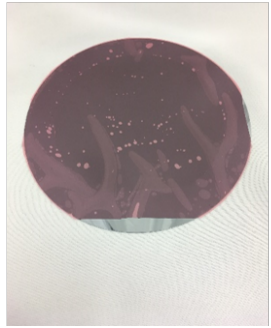 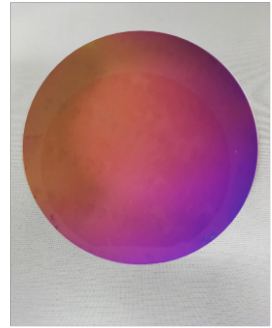

**Figure 2: General procedure of the semidry transfer.**

## 3. Methods

The quality of the transfer was determined by optical contrast of the graphene layer on the target substrate. First, macroscopic analysis was performed with a VHX digital microscope (Keyence) and then, on a microscopic basis, with an OLS4000 laser scanning microscope (Olympus). The macroscopic images were processed with the image processing software PicEd Cora v10.53, and the microscopic images were processed with MATLAB R2018b to determine the degree of coverage by contrast adjustment. These steps were followed for WT1, SDT1, SDT2 and SDT4. The procedure for the optical determination of the degree of coverage is shown in Figure 3.

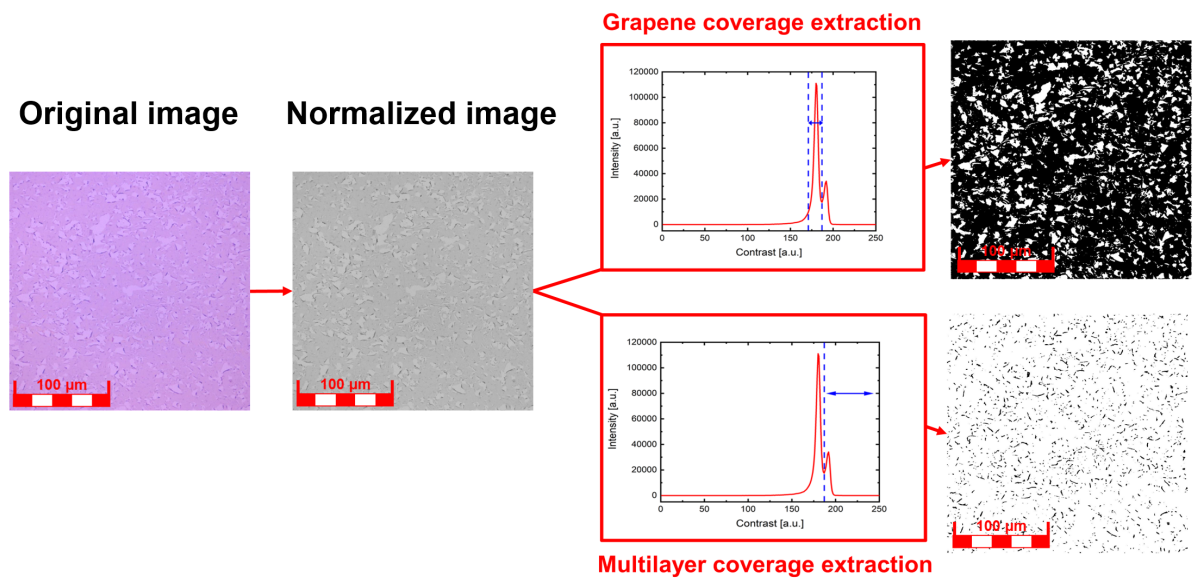

**Figure 3: Determination of the degree of coverage of the graphene layer on the target substrate by optical contrast spectroscopy. Initially, the original image was normalized and converted to a black-and-white image, and then the brightness distribution was determined with MATLAB R2018b. This distribution was used to identify the proportions of substrate, graphene and multilayer graphene. Multilayer in this case describes graphene that is rolled up or wrinkled during the transfer process or initially grown as multilayer graphene.**

Terahertz time-domain spectroscopy (THz-TDS) measurements were performed to determine the effect of the transfer processes on the electrical properties of the transferred graphene. The sheet resistances of the graphene samples WT1 and SDT2 were investigated with a THz near-field inspection system equipped with a photoconductive near-field micro-probe detector

(Protemics TeraCube M2 with TeraSpike TD-800-X-HR-WT) [20,23,24]. From the THz measurements the sheet resistance of the graphene layer on the target substrate can be extracted and thus conclusions can be drawn about the transfer quality. The sheet conductivity ($\sigma_{sh} = 1/R_{sh}$) of the graphene layer $\sigma_{sh}$ can be calculated from the THz data with the Tinkham formula [25]:

$$T(\omega) = \frac{T_{SL}(\omega)}{T_S(\omega)} = \left(1 + \frac{\sigma_{sh}(\omega) \cdot Z_0}{(n+1)}\right)^{-1}, \tag{1}$$

where $T_{SL}$ is the THz transmission through substrate and graphene and $T_S$ is the THz transmission only through substrate. $Z_0$ is the freespace impedance ($Z_0 = 377\ \Omega$) and $n$ is the THz refractive index of the substrate (n = 3.42 for silicon) [25]. The wafer area not covered by the 150mm of transferred graphene was smaller than the substrate (200 mm) and was used as reference for the determination of $T_S$. A detailed overview of the measurement methodology can be found in [20,26]

The sheet resistance of the graphene is sensitive to contamination, defect densities, etc. [27,28]. It was also measured electrically for direct comparison to the THz measurements. WT1 and SDT2 were etched into van der Pauw structures after the THz measurements on the entire wafer to extract reliable electrical data [29].

The metallic residues, in particular Cu contamination from the graphene growth process [30], were measured with the total reflection X-ray fluorescence (TXRF) method on WT2 and SDT3. Here, the copper contamination during wet and semidry transfer on $SiO_2$ was analyzed in comparison to an untreated $SiO_2$ reference wafer.

X-ray photoelectron spectroscopy (XPS) was performed with an ESCALAB 250 Xi XPS system (Thermo Scientific) to determine the degree of polymer-based impurities originating from the transfer processes. As with the TXRF measurements, a reference sample ($SiO_2$) and a wet- and semidry-transferred sample (WT3 and SDT4) were examined, and the carbon bonding types were analyzed. However, the XPS measurement was performed only at a locally restricted area in the wafer center of the sample, as this measurement method does not allow the creation of complete wafer maps like TXRF and therefore does not allow a waferscale quantitative evaluation of the transfer processes.

Raman spectroscopy was carried out with a Horiba Raman system with a laser excitation wavelength of 532 nm (2.33 eV) and a spot size in x and y direction of ~1.05 µm by using a 100x objective with a long working distance focusing lens with 0.21 mm and a numerical aperture of 0.90. A single-mode optical fiber and a spectrometer with a grating of 1,800 lines/mm was used for detection and all measurements were performed at a power density of 5.77 mW/µm$^2$ with linear laser polarization and an integration time of 1 s.

## 4. RESULTS AND DISCUSSION

**Coverage and yield**

The quality of the graphene transfer during wet and semidry transfer was investigated by contrast spectroscopy. For this purpose, the samples were first examined at the macroscopic level (wafer level) to identify different areas of transfer quality and then at the microscopic level in these areas (see methods section). The wet transfer exhibited a microscopic transfer yield of graphene of 97.9 % to 99.2 % distributed over the wafer WT1. A total of seven wafers were prepared for the wet transfer experiments, but four of these were destroyed during transfer (yield = 0%). This is due to the challenge of handling the graphene during the process, where only PMMA remains as a stabilizing support after the copper is etched. For the remaining three, the yield was comparable.

The coverage analysis of the wet transfer (WT1) is shown in **Fehler! Verweisquelle konnte nicht gefunden werden.**. The high transfer quality of wet transfer can be seen in the 4 regions (97.9 % to 99.2 %). The mechanical forces during the transfer process are rather low during wet transfer compared to semidry transfer with its bonding process. We therefore propose that the combination of low mechanical stress during the transfer process and the stabilizing carrier polymer results in the yield close to 100%.

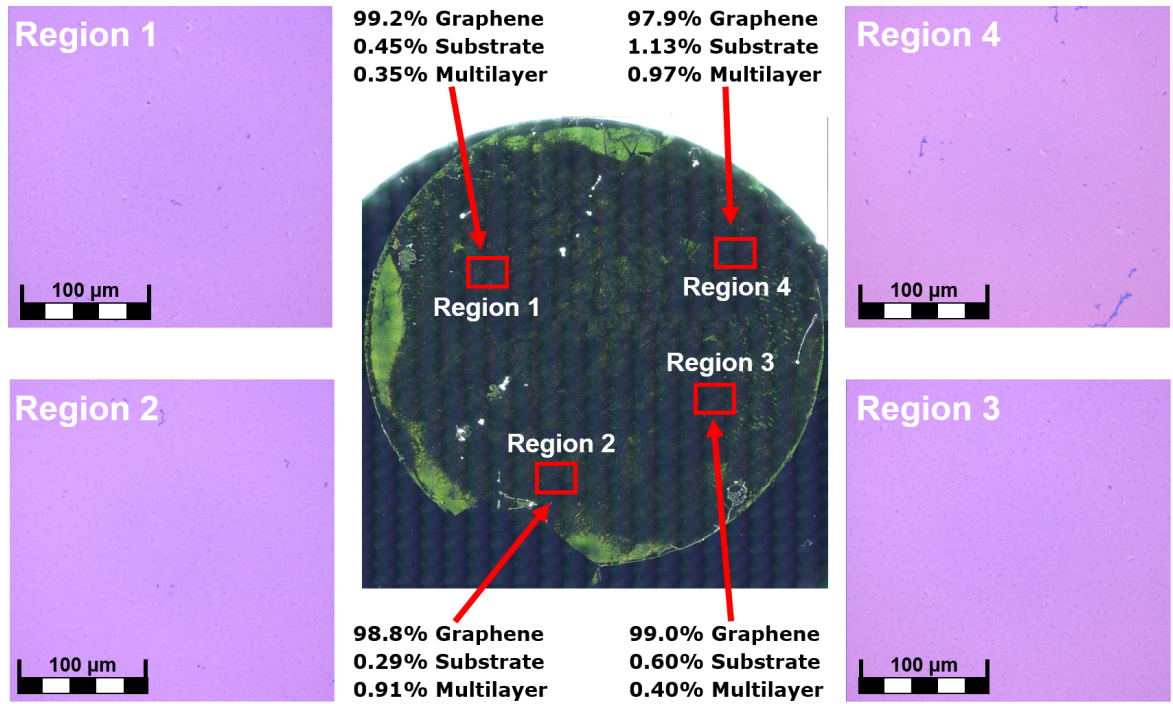

**Figure 4: Contrast image of wet-transferred graphene on the basis of WT1 and four different regions of transfer yield.**

For semidry transfer, the yield varied depending on the pressure of the second bonding step in which graphene was transferred to the target substrate. The resulting yields obtained from four different areas over the wafer in Figure 4 were 28.8 % to 61.3 % at a bonding pressure of 1 N/mm$^2$, 17.2 % to 87.4 % at a bonding pressure of 2 N/mm$^2$ and 71.0 % to 90.8 % at a bonding pressure of 3 N/mm$^2$, respectively. Based on these data, we conclude that the transfer yield depends on the bond pressure applied to the transferred graphene on the target substrate.

Figure 5 shows an example of macroscopic and microscopic coverage determination. Figure 5 a), b) and c) show contrast images of the transferred graphene on SiO$_2$ with bonding pressures of 1 N/mm$^2$ (SDT1), 2 N/mm$^2$ (SDT2) and 3 N/mm$^2$ (SDT4), respectively. The green areas in the contrast images indicate the transferred graphene, and the blue areas indicate areas where no graphene was transferred. The red areas in b) and c) are marked differently because there was no graphene in that area to begin with (wafer flats of the graphene source). Figure 5 d) shows the contrast image from c) in the center, with four adjacent regions from which the graphene transfer and the proportions of substrate, graphene and multilayer (rolled up or wrinkled) graphene were observed on a microscopic basis. These were determined via the contrast analysis of the graphene components described in the Methods section and as shown in Figure 3. The contrast images show that the transferred graphene has areas with lower (region 1) or higher defect densities (region 3). This may result from inhomogeneities of the original graphene on copper foil, from adhesion inhomogeneities caused by the waviness of the TRT, or by contamination particles on the $SiO_2$ surface.

Additionally, the contrast of transferred graphene was much higher for semidry transfer than for wet transfer, which may indicate higher polymeric or metallic contamination for semidry transfer, e.g., residues from the thermal release tape that enhance the optical contrast.

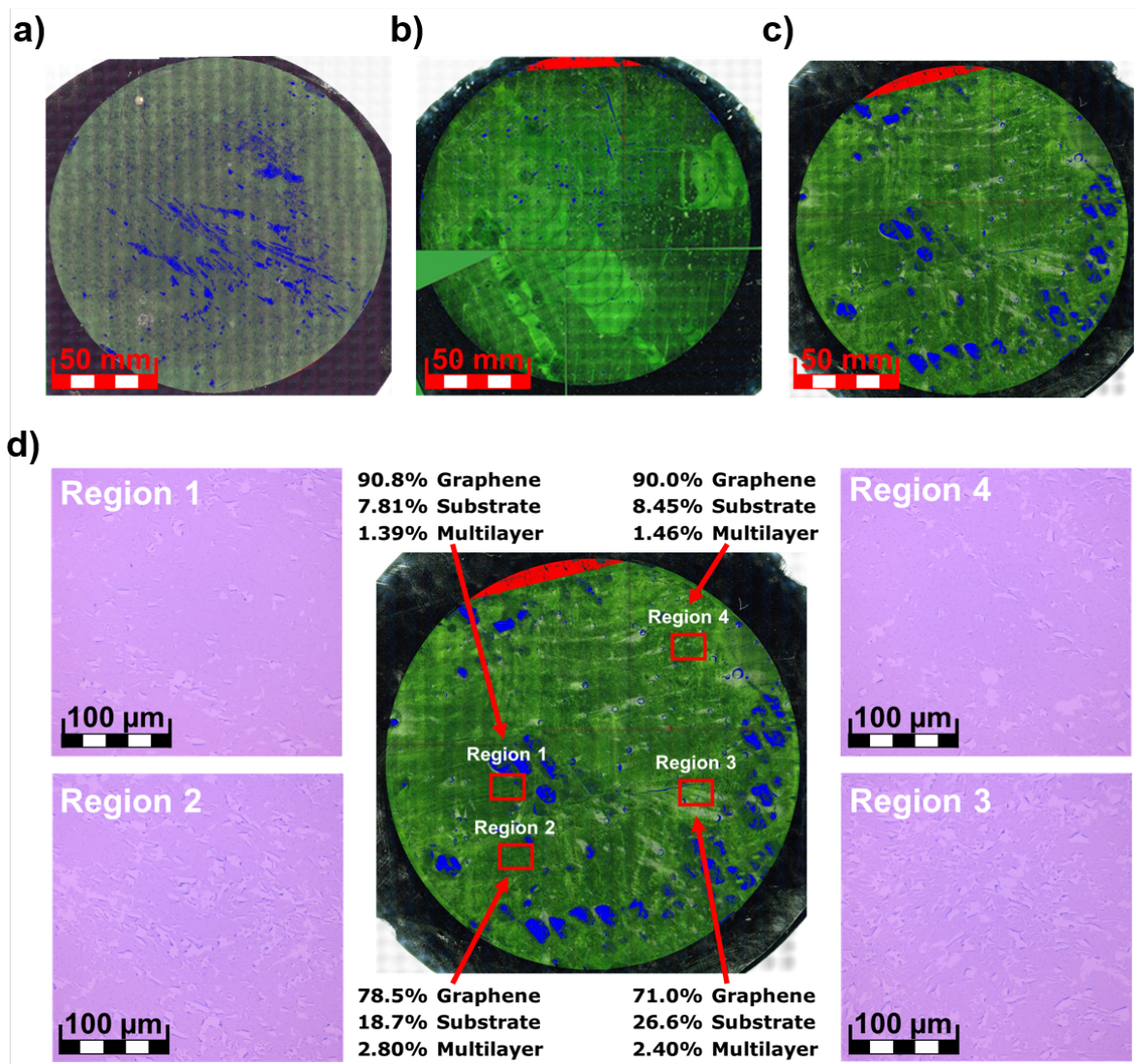

Figure 5: Contrast images of semidry transferred graphene at 1 N/mm$^2$ (SDT1) a), 2 N/mm$^2$ (SDT2) b), and 3 N/mm$^2$ (SDT4) c) and microscopic transfer yield analysis on SDT4 d). Red indicates the flat of the original Cu wafer foil with graphene marked (no contrast), blue indicates areas where no graphene was transferred (no contrast), and green indicates areas where graphene is transferred. In d), four different regions are visible, in which the graphene yield and portion compared to substrate and multilayer regions differ.

**Contamination**

Total reflection X-ray fluorescence (TXRF) and X-ray photoelectron spectroscopy (XPS) measurements were performed to investigate the degree of contamination of wet and semidry transfer (for details see methods section). The TXRF measurements reveal metallic contaminants, which are critical because they lead to increased junction leakage currents and degraded lifetimes and dielectric strengths in integrated circuits [31]. Additionally, at very low concentrations ($10^{10}$-$10^{11}$ atoms/cm$^2$), trace metals can cause a serious threat to silicon devices [32]. Here, special attention was given to copper contamination originating from the growth copper foil, as no media containing obvious metallic components were used during the transfer.

Two separate wafers were prepared via wet and semidry transfer (WT2 and SDT3) for TXRF measurements. Since the contamination class of the TXRF measurement tool had a higher classification (indicating a lower level of allowed contamination) than the tool for contrast determination, no coverage determination was carried out before this experiment. Thus, a one-to-one correlation between metallic contamination and graphene contrast could not be determined for the same wafer. The transfer results under the same conditions used for wet and semidry transfer as used for WT1 and SDT2 (Figure 4 and Figure 5 b) were compared. XPS measurements were carried out to investigate polymeric contamination, but also prohibit same-wafer correlations with optical methods, because they are performed very locally on the wafer. To achieve indirect correlation, SDT4 was used, and an additional wet-transferred wafer (WT3) was prepared, along with a reference wafer (SiO$_2$ on silicon) that was analyzed via TXRF and XPS measurements to determine the level of contamination prior to transfer.

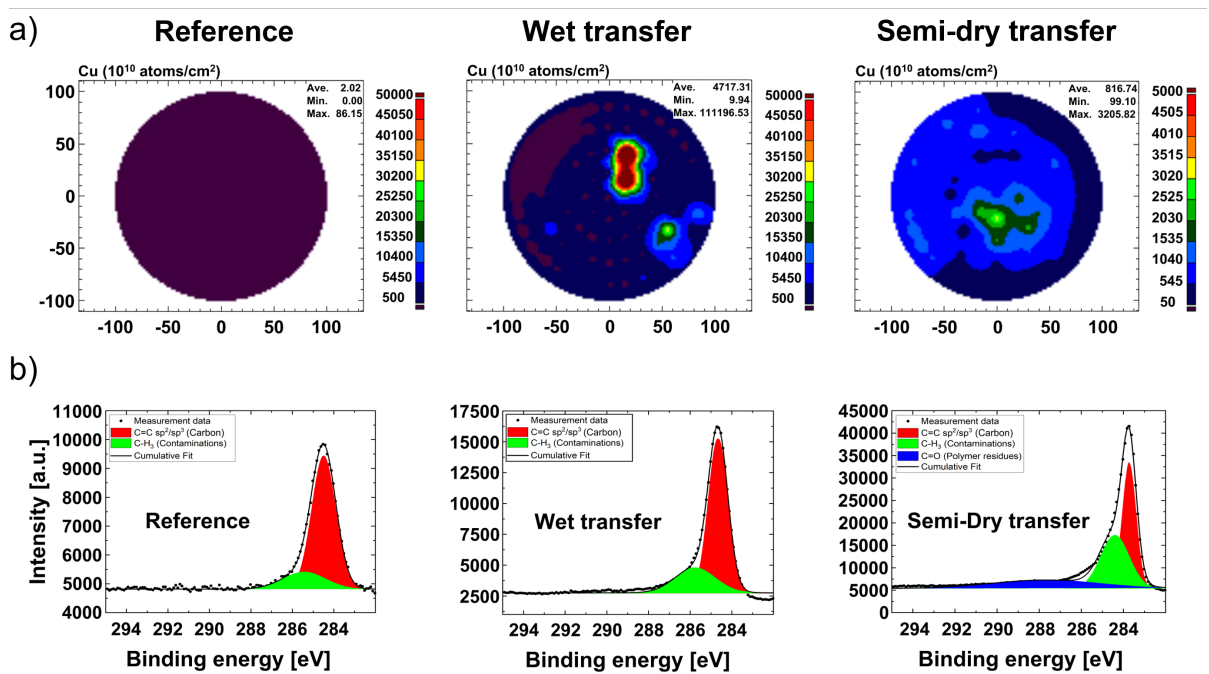

**Figure 6:** Results the TXRF measurements a) and XPS measurements b) for SiO$_2$ (reference) and the wet- and semidry-transferred wafers. a) The TXRF measurements show that the copper contamination during wet transfer is significantly higher on average than that during semidry transfer due to the better handling of the semidry transfer process. The copper contamination on the reference wafer is in the range of 10$^{10}$ atoms/cm$^2$, which is the detection limit of the TXRF system and satisfies the front end of line (FEOL) requirements. b) The XPS measurements show that the degree of carbon contamination, which originates from polymer residues, is lower during wet transfer than during semidry transfer while both are higher compared with that of the reference wafer.

The TXRF data of the reference wafer shows an average copper contamination of 2x10$^{10}$ atoms/cm$^2$, while WT3 showed an average copper contamination of approximately 4.7x10$^{13}$ atoms/cm$^2$ and SDT4 showed an average of approx. 8.2x10$^{12}$ atoms/cm$^2$ (Figure 6a). The lower detection limit of the TXRF system is 10$^{10}$ atoms/cm$^2$, which means that the value determined for the Cu contamination of the reference wafer is an upper estimate. Note that the reference wafers were handled with tweezers, resulting in local contamination with slightly higher than 10$^{10}$ atoms/cm$^2$. The graphene-covered wafers showed high contamination levels, with most of the wet-transferred graphene being slightly less contaminated than the semidry-transferred graphene (the darker the blue color is, the lower the copper contamination). However, there were local areas of wet-transferred graphene with very high copper contamination (e.g., approximately $25 \cdot 10^{13}$ atoms/cm$^2$ and $45 \cdot 10^{13}$ atoms/cm$^2$). There were similar high-contamination spots in semidry-transferred graphene ($25 \cdot 10^{12}$ atoms/cm$^2$), but these had a lower average copper contamination level than those in wet-transferred graphene. We attribute the higher contamination levels after wet transfer to the lack of convection during the Cu etch, because the low stability of the graphene prevented the use of mechanical stirring. Without stirring, areas remain where the copper cannot be completely removed due to local oversaturation of the etching solution with copper. During semidry transfer, the efficient exchange of the etching and purification media was feasible because the carrier wafer sufficiently stabilized the graphene. Nevertheless, the copper contamination in semidry-transferred graphene is still well above typical CMOS specifications, inline with previous works [30] on a sample basis.

XPS measurements show the concentration of sp$^2$/sp$^3$-hybridized carbon species, which originate from graphene, but also from carbon-containing contaminants. The carbon level was lowest in the reference, followed by wet transfer and then semidry transfer (Figure 6). Moreover, the concentration of methyl groups was lower for wet transfer than for semidry transfer. The XPS data also reveal the presence of carbonyl groups after the semidry transfer, which are standard components of polymers. We attribute the higher levels of carbon contamination and polymer residues in the semidry method to the use of the thermal release tape. These carbonyl groups have been shown to influence the charge carrier concentration in graphene through polar C=O$^-$ bonds and can act as scattering centers for the charge carriers in graphene and thus influence the charge carrier mobility and sheet resistance [33]. Generally, the carbon contamination needs to be further reduced by improved cleaning processes, e.g. by increasing the cleaning time in acetone, the thermal annealing time or the annealing temperatures.

**Electrical measurements**

The yield and quality of the graphene transfer methods was also analyzed with respect to the electrical properties of the graphene on the target substrates. For this purpose, THz time-domain spectroscopy and electrical measurements were performed to extract the graphene sheet resistance [20,23,24], which is sensitive to metallic and polymeric (ionic) impurities [34,35]. Details are given in the methods section. Figure 7 shows the THz measurements of WT1 and SDT2 in comparison with the contrast images.

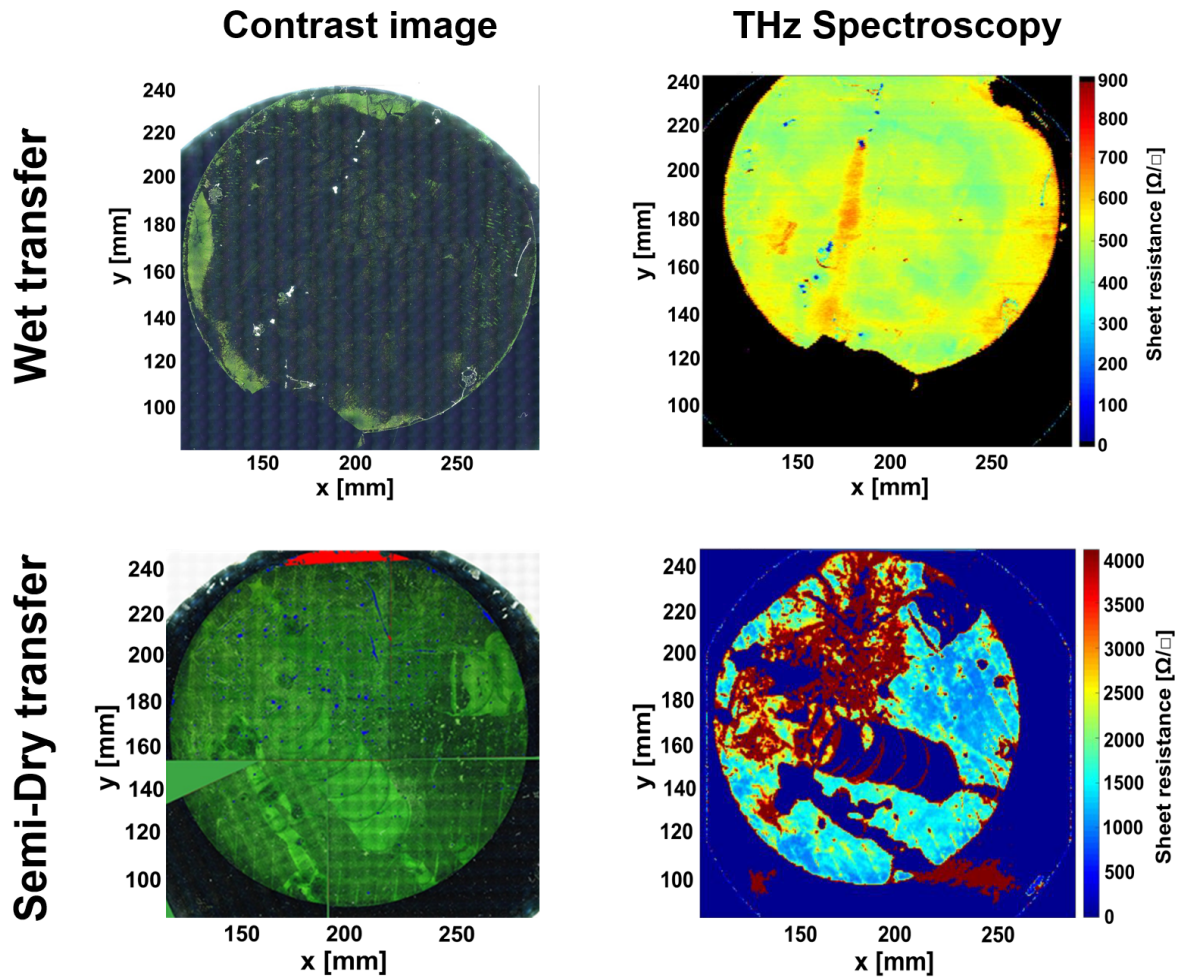

**Figure 7: Comparison of contrast images of wet-transferred (WT1) and semidry-transferred (SDT2) samples on the left side and the extracted sheet resistance from THz measurements on the right side. It can be seen that a lower contrast correlates to a lower sheet resistivity in the THz measurements, which indicates that polymeric and/or metallic contaminants have an influence. The sheet resistance of WT1 is in the range of 450-550 Ω/□, and that of SDT2 is in the range of 1000-1650 Ω/□.**

The THz measurements show a sheet resistance of graphene on $SiO_2$ for WT1 in the range of 450-550 Ω/□, while for SDT2, the range is 1000-1650 Ω/□ over the corresponding wafer. By comparing the electrical sheet resistance values with the optical contrast of the transferred graphene, we observe that the lower the contrast of the graphene is, the lower the sheet resistance for the used material stack and substrate thicknesses. We attribute the higher contrast to higher polymeric content since the metallic impurities have a similar content over the wafers from both transfer methods. Also, no contrast hotspots were observed, although copper contamination hotspots are visible in the TXRF data in Figure 6. This finding supports the assumption that polar $C=O^-$ bonds have a strong influence on the electrical sheet resistance, since they serve as scattering centers on the graphene surface and thus increase the resistance. The fact that the contrast over the entire graphene surface is similarly high for semidry transfer also suggests that these polymeric contaminations originate from the thermal release tape ($C=O^-$ bonds). The increased sheet resistance after semidry transfer suggests that polar $C=O^-$ bonds are responsible for deteriorating the electrical properties of graphene and subsequent device performance.

In addition to the electrical resistances obtained from the THz measurements, both wafers (WT1 and SDT2) were subsequently contacted and lithographically structured to determine their electrical resistivity and charge carrier mobility from four-point measurements. WT1 showed a resistivity of approximately 100-400 $\Omega/\square$ and a charge carrier mobility of 1.000-1.200 $cm^2/V \cdot s$, which are in good agreement with the results of the THz measurements regarding electrical sheet resistance. SDT2, in contrast, showed a resistivity of several $k\Omega/\square$, while the charge carrier mobility could not be determined. We attribute this to macroscopic defects in the graphene film that influence large scale electrical measurements more than local THz spectroscopy. Figure 8 shows the graphene from SDT2 after contacting a) and after etching it into van der Pauw structures b).

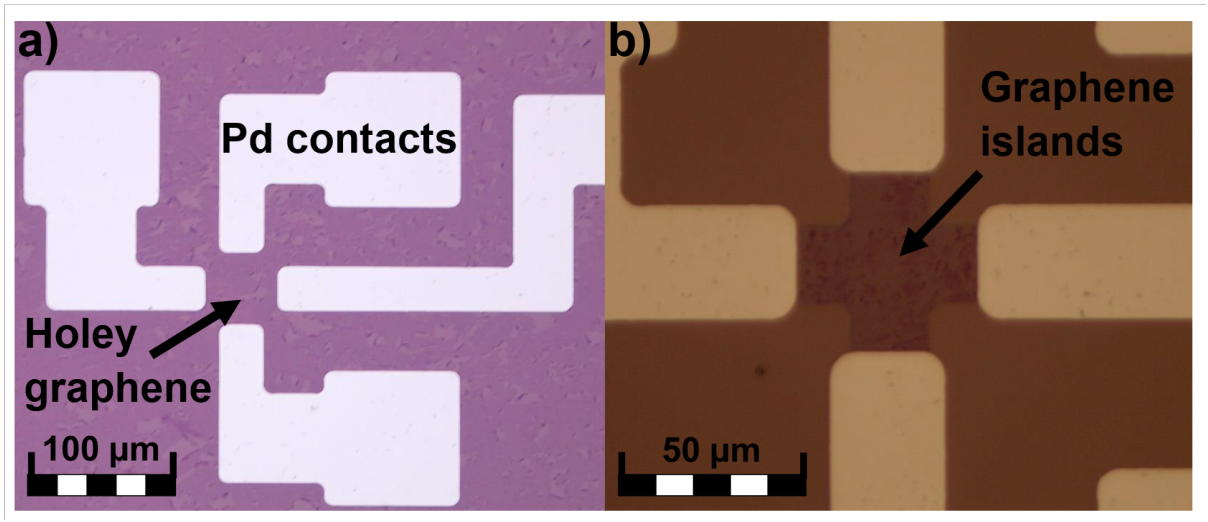

**Figure 8:** Optical images of structured graphene from SDT2 on SiO$_2$ for electrical measurements on two different devices in wafer. a) Graphene shows many defects before patterning due to transfer and contact. b) After patterning, graphene islands are visible on the van der Pauw test structure, which originate from carbon residues during copper etching and from residues of the thermal release tape.

**Fehler! Verweisquelle konnte nicht gefunden werden.** provides a qualitative assessment of the current status of wet and semidry transfer based on this work regarding the different potential process impacts.

**Table 2:** Overview of the current status of wet and semidry transfer regarding handling during transfer, its degree of automation, the transfer yield, metallic and polymeric contamination and the electrical sheet resistance of the transferred graphene.

|  | Wet transfer | Semidry transfer |
|---|---|---|
| **Handling during transfer** | low | high |
| **Degree of automation** | low | high |
| **Yield of transferred graphene** | high | medium |
| **Metallic contamination** | medium | low |
| **Polymeric contamination** | low | medium |
| **Electrical performance** | high | medium |

## 5. CONCLUSIONS

We assessed the feasibility of wet and semidry techniques to transfer graphene onto $SiO_2$ surfaces at the wafer level, as well as the influence of these processes on the contamination levels and electrical performance of the graphene. Wet transfer, often used for chip-scale experiments, poses severe challenges at the wafer level, since the graphene is not stabilized by a carrier substrate, a problem which scales with the graphene area. The quality and yield of graphene transferred with the semidry method increases with increasing bonding force. Nevertheless, the yield and uniformity of the graphene is still higher for wet transfer, if successful, because during semidry transfer, the external forces are higher due to the bonding processes. Metallic contamination is similar for both transfer methods, and beyond the permitted concentrations for silicon CMOS processing. In addition, wet transfer results in local hotspots of copper contamination that we attribute to a lack of mixing of etch and purification media because the low stability of the graphene did not allow mechanical stirring of etch media. The high copper contamination during graphene integration into front-end-of-line processes can in principle be avoided by using alternative growth substrates such as sapphire [36]. The semidry method relies on thermal release tape, which introduces carbon contamination and also influences the electrical properties. This is evident in the higher polar contamination level of the C=O$^-$ bond portions, as seen from XPS measurements. THz spectroscopy showed a lower resistivity for wet-transferred graphene (450-550 Ω/□) than for semidry-transferred graphene (1000-1650 Ω/□). Four-point measurements confirmed these results (wet-transfer: 100-400 Ω/□; semidry-transfer: several kΩ). In summary, wet transfer leads to higher yield, lower carbon contamination and better electrical performance of the graphene compared to semidry transfer. However, wet transfer has a major disadvantage in handling graphene after copper etching since even minimal mechanical stress in the graphene layer destroy it. In contrast, the carrier wafers used during semidry transfer facilitates handling and thus the automation of the transfer process. Although both transfer techniques have advantages and disadvantages, which may be overcome through more optimized processes and tools, we believe that wafer bonding will ultimately have a decisive advantage in applicability and scalability.

**AUTHOR INFORMATION**

**Corresponding Authors**

E-mail: Univ.-Prof. Dr.-Ing. Max Christian Lemme, max.lemme@eld.rwth-aachen.de

The manuscript was written through contributions of all authors. All authors have given approval to the final version of the manuscript.

ACKNOWLEDGMENT

This work has received funding from the German Ministry of Education and Research (BMBF) under the project GIMMIK (03XP0210) and from the European Union's Horizon 2020 research and innovation program under grant agreements 881603 (Graphene Flagship Core3), 785219 (Graphene Flagship Core2) and 952792 (2D Experimental Pilot Line).